\documentclass[twocolumn,preprintnumbers,amsmath,amssymb,superscriptaddress]{revtex4-2}
\UseRawInputEncoding

\usepackage{latexsym,amssymb,amsthm,amsmath,epsfig,braket}
\usepackage[left=2.00cm, right=2.00cm, top=2.00cm, bottom=2.00cm]{geometry}
\usepackage{xcolor}
\usepackage{svg}
\usepackage[colorlinks, citecolor={blue!80!black}, urlcolor={blue!80!black}, linkcolor={red!50!black}]{hyperref}
\usepackage{float}
\usepackage{hyperref}
\hypersetup{
    citecolor=red,
    colorlinks=true,
    linkcolor=blue,
    filecolor=blue,      
    urlcolor=blue,
}

\begin{document}

\title{Floquet-driven thermal transport in topological Haldane lattice systems}

\author{Imtiaz Khan}
\affiliation{Department of Physics, Zhejiang Normal University, Jinhua 321004, P. R. China}
\affiliation{ Zhejiang Institute of Photoelectronics $\&$ Zhejiang Institute for Advanced Light Source, Zhejiang Normal University, Jinhua, 321004, P. R. China}
\author{Muzamil Shah}
\email{muzamil@qau.edu.pk}
\affiliation{Department of Physics, Quaid-I Azam University Islamabad, 45320, Pakistan}
\affiliation{Research Center of Astrophysics and Cosmology, Khazar University, Baku, AZ1096, 41 Mehseti Street, Azerbaijan}
\author{Ambreen Uzair}
\affiliation{National Centre for physics, Islamabad, 45320, Pakistan}
\author{Reza Asgari}
\affiliation{Department of Physics, Zhejiang Normal University, Jinhua 321004, P. R. China}
\affiliation{School of Physics, Institute for Research in Fundamental Sciences (IPM), Tehran 19395-5531, Iran}

\author{Gao Xianlong}
\email{gaoxl@zjnu.edu.cn}
\affiliation{Department of Physics, Zhejiang Normal University, Jinhua 321004, P. R. China}

\begin{abstract}
In this paper, we employ a modified Haldane lattice model to investigate the light-driven, spin-and valley-dependent anomalous Nernst effect in two-dimensional hexagonal topological systems. We demonstrate that two-dimensional buckled materials exhibit a hierarchy of electrically and optically tunable topological phases when subjected to off-resonant circularly polarized light in the presence of intrinsic spin-orbit coupling and a staggered sublattice potential. Within a Berry-curvature-driven transport framework, we systematically analyze charge-, spin-, and valley-resolved anomalous Nernst responses and identify their correspondence with distinct topological regimes. A finite charge Nernst conductivity arises under optical driving combined with spin-orbit coupling, whereas the generation of a pure valley Nernst current requires the simultaneous presence of sublattice asymmetry and off-resonant light. Substrate-induced inversion asymmetry further enables thermally driven valley currents with tunable magnitude and sign. We find that single-spin and single-valley Nernst responses occur in selected insulating and metallic phases, while the valley Nernst signal is suppressed in spin-polarized and anomalous quantum Hall phases. Extending our analysis to monolayer MoS$_2$, we show that strong spin–orbit coupling and broken inversion symmetry allow fully spin- and valley-polarized Nernst currents over a broad energy window. The temperature dependence of the Nernst response exhibits characteristic signatures of topological phase transitions, establishing the anomalous Nernst effect as a sensitive probe of field-engineered band topology in two-dimensional Dirac materials.

\end{abstract}

\maketitle
\newpage
\section{Introduction}
\label{int}

Two-dimensional (2D) crystalline materials~\cite{Novoselov2004,novoselov2004electric} attracted significant interest due to their nontrivial electronic properties and band topology. These systems span a wide range of behaviors, from semiconducting to superconducting, including buckled Xenes such as silicene, germanene, and stanene, as well as transition metal dichalcogenides (TMDs) like $\mathrm{MoS_{2}}$ and $\mathrm{WSe_{2}}$~\cite{ezawa2012topological,ezawa2013spin,Bampoulis202023quantum,CHOI2017116} . Their tunable band structures, strong spin–orbit coupling (SOC)~\cite{PhysRevB.100.125422}, and valley-dependent physics~\cite{schaibley2016valleytronics}, make them promising platforms for exploring topological transport and developing next-generation spintronic and valleytronic devices~\cite{pesin2012spin,schaibley2016valleytronics}.

The electronic properties of hexagonal lattices are governed by time-reversal ($\mathcal{T}$) and inversion ($\mathcal{I}$) symmetries~\cite{du2021engineering}. Breaking these symmetries alters the Berry curvature and can induce topological phase transitions~\cite{RevModPhys.82.1959,ren2016topological}. A key example is the Haldane model~\cite{PhysRevLett.61.2015}, where complex next-nearest-neighbor hopping breaks $\mathcal{T}$ symmetry and produces a quantum Hall-like state without a magnetic field. In real 2D materials, effects such as intrinsic SOC, sublattice potentials, and external fields further enrich and enable tunable topological phases.

The anomalous Nernst effect (ANE)~\cite{mizuguchi2019energy}, a transverse electric current generated by a longitudinal temperature gradient without a magnetic field, has emerged as a key probe of band topology~\cite{RevModPhys.82.1959}. Related phenomena, including spin, valley, and nonlinear Nernst effects~\cite{PhysRevB.78.045302,meyer2017observation,PhysRevLett.115.246601,dau2019valley,PhysRevB.98.060402,PhysRevB.99.201410}, arise from Berry curvature acting as an effective magnetic field in systems with broken $\mathcal{T}$ or $\mathcal{I}$ symmetry. Furthermore, nonlinear anomalous Nernst responses arising from higher-order Berry curvature effects have been reported in Dirac systems with trigonal warping~\cite{wu2021nonlinear}. The present work focuses on the linear anomalous Nernst effect governed by the Berry curvature of massive Dirac bands, where the response is directly linked to symmetry breaking and Floquet-induced topological phase transitions. In Dirac materials, spin–orbit coupling, exchange fields, or inversion asymmetry can induce valley-contrasting Berry curvature, leading to strong Nernst responses~\cite{zhou2015anomalous}. Consequently, 2D materials such as graphene, silicene, and TMDs are promising platforms for spin- and valley-caloritronics~\cite{zhu2013anomalous,niu2022highly,vargiamidis2020berry,sharma2018tunable,hajati2023magnetoelectric,tahir2014tunable}. While the anomalous Nernst effect in graphene-based systems with broken time-reversal symmetry has been studied in ~\cite{zhu2013anomalous}, their analysis was restricted to strain-induced modifications without Floquet driving and did not address spin- or valley-resolved responses. In contrast, our work incorporates periodic driving and spin-orbit coupling, enabling a unified description of charge, spin, and valley Nernst effects across multiple topological regimes.

Periodic driving provides an additional route to control topological transport~\cite{bao2022light}. Within Floquet theory, off-resonant circularly polarized light can modify Dirac band structures by inducing mass terms and topological gaps~\cite{sambe1973steady,shirley1965solution,eckardt2015high,goldman2014periodically,shah2024topological}. This approach enables symmetry breaking and Berry curvature control, allowing optical tuning of topological phases and transport. Combined with SOC and sublattice asymmetry, it also permits selective control of spin and valley degrees of freedom, offering new possibilities for tunable thermoelectric responses in 2D systems~\cite{dai20242d}. In contrast to the Floquet topological analysis of silicene in ~\cite{ezawa2013photoinduced}, where the focus was placed on photoinduced phase transitions and Chern-number classification, our work establishes a direct connection between Floquet-engineered band topology and thermoelectric transport by explicitly computing charge-, spin-, and valley-resolved anomalous Nernst responses across different phases.

Recent studies on Floquet-engineered topological phases in materials such as jacutingaite~\cite{alipourzadeh2023photoinduced} have demonstrated the emergence of photoinduced quantum Hall states and associated transport signatures. However, these works primarily focus on specific material realizations and do not provide a systematic decomposition of thermoelectric responses. In contrast, our approach employs a generalized modified Haldane framework to establish a comprehensive mapping between topological phases and charge-, spin-, and valley-resolved anomalous Nernst effects. Motivated by recent advances, we study the spin- and valley-dependent anomalous Nernst effect in Floquet-driven hexagonal systems within a modified Haldane framework. We focus on buckled Dirac materials such as germanene and extend the analysis to monolayer $\mathrm{MoS_{2}}$, where strong SOC and broken inversion symmetry yield pronounced spin–valley coupling. Furthermore, prior works have treated Floquet topological phases and anomalous Nernst effects separately~\cite{PhysRevLett.97.026603,PhysRevB.79.081406,PhysRevB.102.014307,mak2014valley}, a unified understanding of how Floquet engineering modifies spin- and valley-resolved thermoelectric transport across topological phases remains lacking. Here, we provide such a framework by mapping topological phases to charge-, spin-, and valley-resolved Nernst responses.
We show that optical driving is key to activating and tuning thermoelectric behavior across different regimes. In particular, a pure valley Nernst effect emerges only when inversion symmetry breaking and Floquet driving coexist. The photoinduced Floquet mass acts as a tunable parameter controlling the sign and magnitude of the response, offering a route to engineer spin- and valley-caloritronic effects in 2D Dirac materials.

\section{Model Hamiltonian, Floquet Formalism, and Anomalous Nernst Response}
\label{sec:method}

We describe the low-energy electronic properties of the honeycomb-lattice system within a generalized modified-Haldane Model  framework, often referred to as the modified-Haldane Model~\cite{haldane1988model,vanderbilt2018berry,pratama2020circular}. The corresponding tight-binding Hamiltonian on a 2D honeycomb lattice in this framework reads as
\begin{equation}
\hat{\mathcal{H}} = 
- t_1 \sum_{\langle i,j \rangle} a_i^\dagger a_j
+ t_2 \sum_{\langle\langle i,j \rangle\rangle} 
e^{i \nu_{ij} \phi} a_i^\dagger a_j
+ \mathcal{M} \sum_i \chi_i a_i^\dagger a_i ,
\label{eq:haldane_tb}
\end{equation}
where $a_i^\dagger$ ($a_i$) creates (annihilates) an electron at lattice site $i$. The parameters $t_1$ and $t_2$ denote nearest-neighbor (NN) and next-nearest-neighbor (NNN) hopping amplitudes, respectively. The phase factor $\nu_{ij}=\pm1$ distinguishes clockwise and counterclockwise NNN hopping paths, and the associated phase $\phi$ explicitly breaks time-reversal symmetry. The last term introduces a staggered sublattice potential, with $\chi_i=+1$ ($-1$) for sublattice $A$ ($B$), which breaks inversion symmetry and opens a mass gap $2\mathcal{M}$ at the Dirac points.

The modified Haldane model represents a minimal theoretical framework for describing topological band structures in honeycomb lattices. It extends the original Haldane model by including additional symmetry-breaking terms that are relevant for realistic Dirac materials. In particular, three key ingredients determine the band topology: (i) nearest-neighbor hopping that generates the Dirac cones, (ii) complex next-nearest-neighbor hopping which breaks time-reversal symmetry and produces a topological mass gap, and (iii) a staggered sublattice potential that breaks inversion symmetry by introducing different on-site energies on the two sublattices. The competition between these terms controls the Berry curvature distribution and determines whether the system realizes trivial insulating, quantum spin Hall, or Chern insulating phases. Expanding the Hamiltonian around the inequivalent valleys $K$ and $K'$ and retaining terms linear in momentum, one obtains an effective low-energy description in terms of a spin-resolved massive Dirac Hamiltonian~\cite{ezawa2013photoinduced,ezawa2015monolayer,alipourzadeh2023photoinduced,shah2025quantum}. For valley index $\tau=\pm1$ and spin projection $s=\pm1$, the Hamiltonian takes the form
\begin{equation}
\hat{\mathcal{H}}(\mathbf{k}) =
\hbar v_F \left( \tau k_x \sigma_x + k_y \sigma_y \right)
-\tau s \lambda_{\mathrm{so}} \sigma_z
+ \Delta_{\tau s}\sigma_z
- \lambda \mathbb{I},
\label{eq:dirac}
\end{equation}
where $\sigma_i$ are Pauli matrices acting in sublattice space and $\mathbb{I}$ is the $2\times2$ identity matrix. The Fermi velocity is given by $v_F=\sqrt{3} a t/(2\hbar)\approx 10^6~\mathrm{m/s}$, with lattice constant $a=2.46~\text{\AA}$ and NN hopping amplitude $t=2.97~\mathrm{eV}$. The intrinsic SOC strength is taken as $\lambda_{\mathrm{so}}=43~\mathrm{meV}$. The parameters $\lambda$ and $\Delta_{\tau s}$ encode the effects of the complex NNN hopping and staggered sublattice potential,
\begin{align}
\lambda_{\mathrm{so}} &= 3 t_2 \cos\phi_s, \label{eq:lambda_def} \\
\Delta_{\tau s} &= M - \tau\,3\sqrt{3}\,t_2 \sin\phi_s,
\label{eq:delta_def}
\end{align}
where $\phi_s$ denotes the spin-dependent modified-Haldane phase. This formulation allows access to a wide range of topological phases, including quantum spin Hall insulators, Chern insulators, and valley-polarized metallic states.

To dynamically engineer the band topology, we subject the system to circularly polarized light and employ the Floquet formalism to treat the resulting time-periodic Hamiltonian~\cite{sambe1973steady,shirley1965solution,eckardt2015high,goldman2014periodically}. The electromagnetic field is introduced through the time-dependent vector potential
\begin{equation}
\mathbf{A}(t) = A_0\left(\gamma\sin\omega_0 t, \cos\omega_0 t\right),
\end{equation}
where $\gamma=\pm1$ labels the helicity of the light, $A_0=E_0/\omega_0$ is the field amplitude, and $\omega_0$ is the driving frequency. The total Hamiltonian becomes
\begin{equation}
\hat{H}(t)=\hat{H}_0+\hat{V}(t),
\end{equation}
with $\hat{H}_0$ given by Eq.~\eqref{eq:dirac} and the light matter interaction
\begin{equation}
\hat{V}(t)=\frac{e v_F}{\hbar}
\left[\tau A_x(t)\sigma_x + A_y(t)\sigma_y\right].
\end{equation}

In the off-resonant regime, defined by $\hbar\omega_0 \gg e v_F A_0$, the system is governed by an effective static Floquet Hamiltonian obtained via a high-frequency expansion~\cite{kitagawa2011transport},
\begin{equation}
\hat{H}_{\mathrm{eff}}=
\hat{H}_0+\frac{[\hat{H}_{-1},\hat{H}_{+1}]}{\hbar\omega_0}
+\mathcal{O}(\omega_0^{-2}),
\label{eq:Heff}
\end{equation}
where $\hat{H}_{\pm 1}$ are the first-order Fourier harmonics of the time-dependent Hamiltonian, that the central sideband has the most significant effect,
\begin{equation}
    \hat{H}_{\pm n} = \frac{\omega_0}{2\pi} 
    \int_0^{2\pi/\omega_0} dt\, \hat{H}(t) e^{\pm i n \omega_0 t}.
\end{equation}
and we neglect the higher order $H_{n}$ for $n>1$. The effective Hamiltonian in Eq.   (\ref{eq:Heff}) is obtained using the high-frequency Floquet-Magnus expansion, which is valid in the off-resonant regime where the photon energy significantly exceeds the electronic bandwidth of the system. For honeycomb Dirac materials, the bandwidth is on the order of a few electron volts ($t \approx 2-3$ eV), implying the condition $\hbar\omega \gg t$ or interaction scales. Under this condition, real photon absorption is suppressed, and the driving field modifies the band structure only through virtual processes. In periodically driven systems, heating can, in principle, destabilize Floquet topological phases. However, in the off-resonant limit, the heating rate is strongly reduced, allowing the system to reach a long-lived Floquet prethermal \cite{Machado2017ExponentiallySH, ho2023quantum} state where the effective Hamiltonian description remains valid. Finite temperature primarily broadens the Fermi distribution \cite{Cowan1957ExtensionOT} but does not destroy the topological band structure provided $kBT \ll \hbar\omega$. This procedure generates a light-induced mass correction
\begin{equation}
\lambda_\omega=\gamma\frac{(e v_F A_0)^2}{\hbar\omega_0},
\label{eq:lambdaomega}
\end{equation}
which depends explicitly on the light polarization and driving frequency. The resulting effective Hamiltonian reads
\begin{equation}
\hat{H}^{\mathrm{eff}}_{\tau s \gamma}
=
-\lambda
+\hbar v_F(\tau k_x\sigma_x+k_y\sigma_y)
+\Delta_{\mathrm{tot}}\sigma_z.
\end{equation}
where $\Delta_{\mathrm{tot}}=\Delta_{\tau s}-\tau s\lambda_{\mathrm{so}}+\tau\lambda_\omega$. The corresponding quasi-energy spectrum is
\begin{equation}
E^{\eta}_{\tau s \gamma}(\mathbf{k})
=
-\lambda
+\eta\sqrt{(\hbar v_F k)^2
+\Delta_{\mathrm{tot}}^2},
\end{equation}
where $\eta=\pm1$ labels the conduction and valence bands.

The Berry curvature associated with band $\eta$ plays a central role in transverse transport phenomena. Using the standard definition in momentum space, we obtain the analytical expression~\cite{PhysRevB.109.235418}
\begin{equation}
\Omega^{\eta}_{\tau s \gamma}(k)
=
-\eta\,\tau\,
\frac{\Delta_{\mathrm{tot}}(\hbar v_F)^2}
{2\left[(\hbar v_F k)^2+\Delta_{\mathrm{tot}}^2\right]^{3/2}},
\label{eq:berrycurv}
\end{equation}
which acts as an effective magnetic field in reciprocal space. The integral of the Berry curvature over the Brillouin zone defines topological invariants, including charge, spin, valley, and spin valley Chern numbers, which classify the underlying topological phases. The transverse transport coefficients, such as the anomalous Hall and Nernst responses, are directly determined by the Chern number. The Berry curvature controls the local geometric features of the electronic bands. 

The magnitude and momentum width of the Berry curvature peak are controlled by the effective Dirac mass $\Delta_{tot}$. A smaller mass gap produces a sharper Berry-curvature distribution near the Dirac points, while a larger gap broadens the distribution in momentum space. Experimentally, $\Delta_{tot}$ can be engineered through three independent mechanisms: (i) tuning the perpendicular electric field that controls the staggered potential M, (ii) modifying the intrinsic SOC via substrate engineering or heavier Xene materials, and (iii) adjusting the amplitude and helicity of the optical driving field which determines the Floquet mass $\lambda_\omega$. Optimizing these parameters allows one to maximize the Berry-curvature-weighted entropy contribution that governs the anomalous Nernst response.

The topological phases of the system are characterized by Berry-curvature-derived invariants. The charge, spin, and valley Chern numbers can be expressed in terms of the Berry curvature $\Omega_{\tau s}(\mathbf{k})$ associated with electrons of valley index $\tau = \pm 1$ and spin index $s = \pm 1$. These quantities are defined as follows
\begin{equation}
C = \frac{1}{2\pi}\sum_{\tau,s}\int_{\mathrm{BZ}} \Omega_{\tau s}(\mathbf{k})\, d^{2}k ,
\end{equation}
\begin{equation}
C_{s} = \frac{1}{2\pi}\sum_{\tau,s} s \int_{\mathrm{BZ}} \Omega_{\tau s}(\mathbf{k})\, d^{2}k ,
\end{equation}
\begin{equation}
C_{v} = \frac{1}{2\pi}\sum_{\tau,s} \tau \int_{\mathrm{BZ}} \Omega_{\tau s}(\mathbf{k})\, d^{2}k .
\end{equation}
Here $C$ denotes the total (charge) Chern number, which determines the quantized Hall conductivity of the system. The quantity $C_s$ represents the spin Chern number, characterizing the spin Hall topology, while $C_v$ corresponds to the valley Chern number that captures the imbalance of Berry curvature between the inequivalent $K$ and $K'$ valleys of the hexagonal Brillouin zone. The integrals are evaluated over the first Brillouin zone, and the Berry curvature $\Omega_{\tau s}(\mathbf{k})$ encodes the geometric properties of the Bloch bands in momentum space. The semiclassical Berry-curvature framework, which has been frequently used in anomalous thermoelectric transport characteristics, is identical to the Kubo formalism in the weak-scattering limit. Within semiclassical transport theory, a longitudinal temperature gradient $\nabla T$ induces a transverse current due to Berry curvature effects. The spin- and valley-resolved thermoelectric current is given by~\cite{tamang2023probing,wu2021nonlinear,you2022anomalous}
\begin{widetext}
\begin{equation}
\mathbf{J}_{\tau s \gamma}
=
-\frac{\nabla T}{T}
\times
\sum_{\eta}
\frac{e}{\hbar}
\int\frac{d^2k}{(2\pi)^2}
\Omega^{\eta}_{\tau s \gamma}(k)
\left[
(E^{\eta}_{\tau s}-E_F)f^{\eta}_{\tau s}
+k_BT\ln\!\left(1+e^{-(E^{\eta}_{\tau s}-E_F)/k_BT}\right)
\right].
\label{eq:current}
\end{equation}
\end{widetext}
Here $f^{\eta}_{\tau s}$ is the Fermi Dirac distribution function. The anomalous Nernst coefficient $N_{\tau s \gamma}$ follows from $J_y=N(-\nabla_x T)$ and can be written as
\begin{equation}
N_{\tau s \gamma}
=
\frac{e k_B}{\hbar}
\sum_{\eta}
\int\frac{d^2k}{(2\pi)^2}
\Omega^{\eta}_{\tau s \gamma}(k)
S^{\eta}_{\tau s}(k),
\label{eq:nernst}
\end{equation}
where
\begin{equation}
S^{\eta}_{\tau s}(k)
=
-f\ln f-(1-f)\ln(1-f)
\end{equation}
is the entropy density, sharply localized near the Fermi level.

Finally, the experimentally relevant charge, spin, and valley Nernst conductivities are defined as
\begin{equation}
N_c=\sum_{\tau,s}N_{\tau s},\quad
N_s=\sum_{\tau,s}s\,N_{\tau s},\quad
N_v=\sum_{\tau,s}\tau\,N_{\tau s}.
\end{equation}
While the charge Nernst response can be directly measured electrically, the spin and valley Nernst currents can be detected using inverse spin or inverse valley Hall measurements in multiterminal geometries~\cite{jiang2013generation,yamamoto2015valley,zhang2022valley}.

\section{Temperature Dependence of the Anomalous Nernst Effect}

To elucidate the temperature dependence of the anomalous Nernst effect (ANE) analytically, we invoke the low-temperature Mott relation connecting the spin-valley-resolved anomalous Nernst and Hall conductivities~\cite{RevModPhys.82.1959},
\begin{equation}
N_{\tau s}
= \frac{\pi^2 k_B^2 T}{3e}\,
\frac{d\sigma_{\tau s}(E_F)}{dE_F},
\label{eq:mott}
\end{equation}
where the corresponding Hall conductivity is given by
\begin{equation}
\sigma_{\tau s}
= \frac{e^2}{\hbar}
\sum_{\eta}\!\!\int\!\frac{d^2 k}{(2\pi)^2}\,
\Omega^{\eta}_{\tau s}(\mathbf{k})\,f^{\eta}_{\tau s}(\mathbf{k}).
\label{eq:hall}
\end{equation}
For $E_F > |\Delta_{\mathrm{total}}|$, i.e., when the Fermi level lies within the conduction band, Eq.~(\ref{eq:hall}) reduces to the compact form
\begin{equation}
\sigma_{\tau s}
= -\frac{e^2}{h}\,
\frac{\tau\,\Delta_{\mathrm{total}}}{4E_F},
\label{eq:sigma_simplified}
\end{equation}
with $E_F=\sqrt{(\hbar v_F k_F)^2+\Delta_{\mathrm{total}}^2}$.  
Substitution into Eq.~(\ref{eq:mott}) yields the analytic low-temperature expression
\begin{equation}
N_{\tau s}
= \frac{\pi^2}{12}\,\frac{e k_B^2 T}{h}\,
\frac{\tau\,\Delta_{\mathrm{total}}}{E_F^2}.
\label{eq:N_lowT}
\end{equation}

\begin{figure*}[t!]
	\centering
    \includegraphics[width=8.0cm, height=9.0cm]{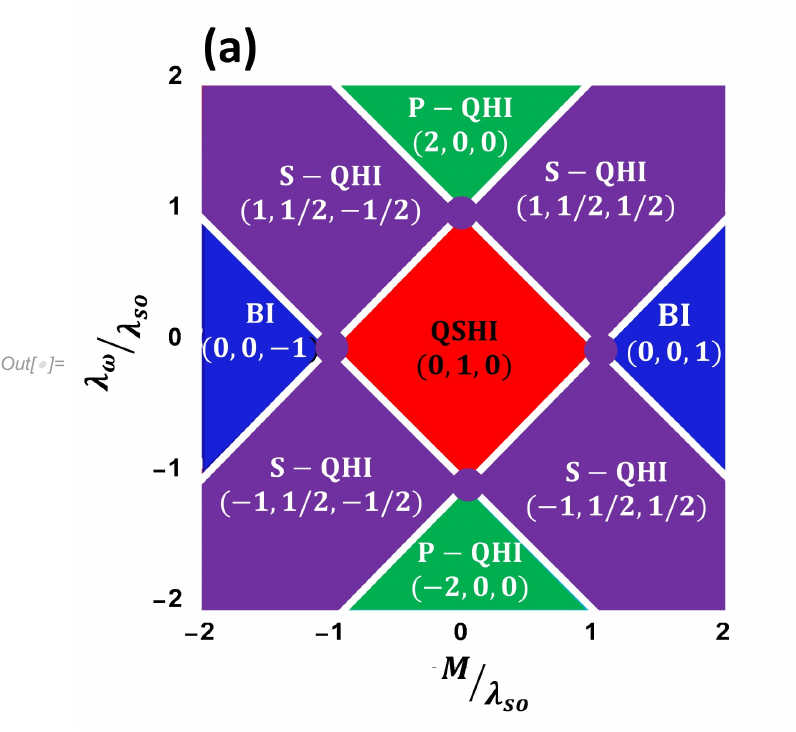}
	\centering
	\includegraphics[width=9.0cm, height=8.5cm]{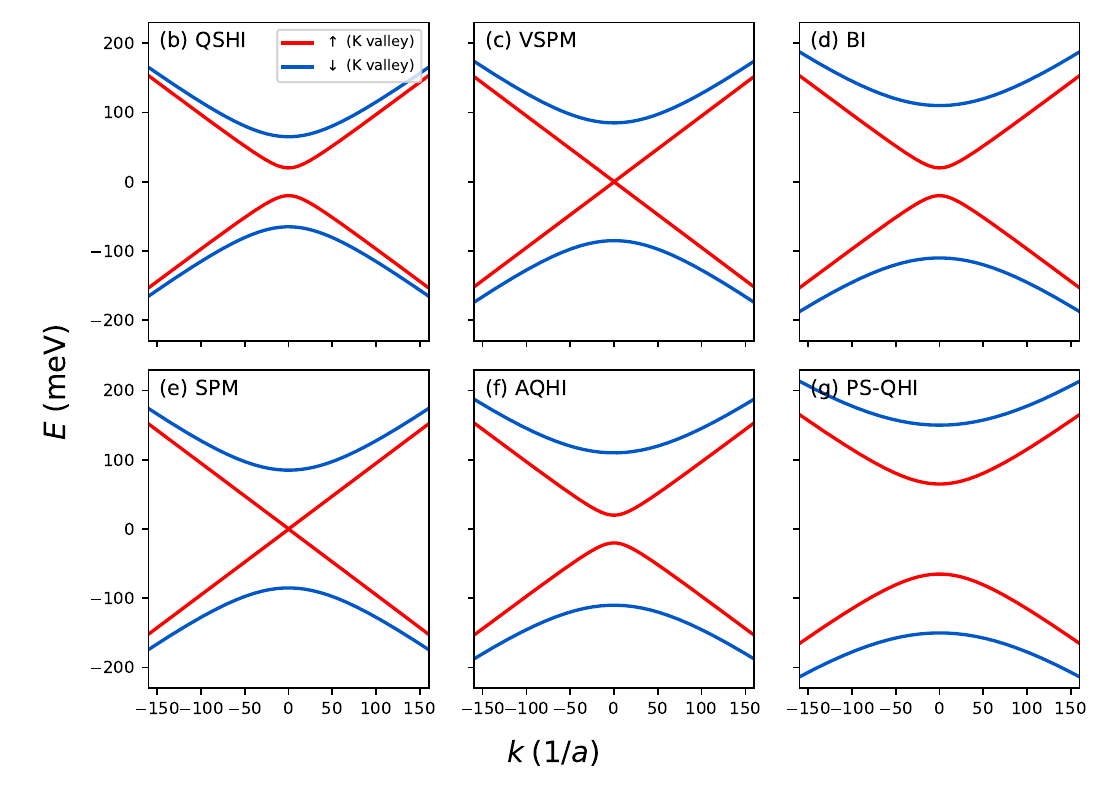}
	\caption{The phase diagram of a monolayer crystal as a function of $M/\lambda_{\textit{so}}$ and $\lambda_{\omega}/\lambda_{\textit{so}}$ in the left panel (a). The distinct electronic phases are labeled by different colors and are indexed by the total, spin, and valley Chern numbers ($\mathcal{C}$, $\mathcal{C}_{s}$ and $\mathcal{C}_{v}$). Band structure of the buckled Xene materials at the $K$ valley for different topological regimes is shown in the right panel. (b) QSHI ($M=0.5\lambda_{\text{so}}$, $\lambda_{\omega}=0$), (c) VSPM  ($M=\lambda_{\text{so}}$, $\lambda_{\omega}=0$), (d) BI ($M=1.5\lambda_{\text{so}}$,  $\lambda_{\omega}=0$), (e) SPM ($M=0$,  $\lambda_{\omega}=\lambda_{\text{so}}$), (f) AQHI ($M=0$, $\lambda_{\omega}=1.5\lambda_{\text{so}}$) and (g) PS-QHI ($M=\lambda_{so}$,  $\lambda_{\omega}=1.5\lambda_{\text{so}}$) respectively. The corresponding phases of the band structures are indicated in the phase diagram. The blue (magenta) curves are for spin-up (down).}
	\label{Energy}
\end{figure*}

\begin{figure*}[th!]
	\includegraphics[width=1.0\linewidth]{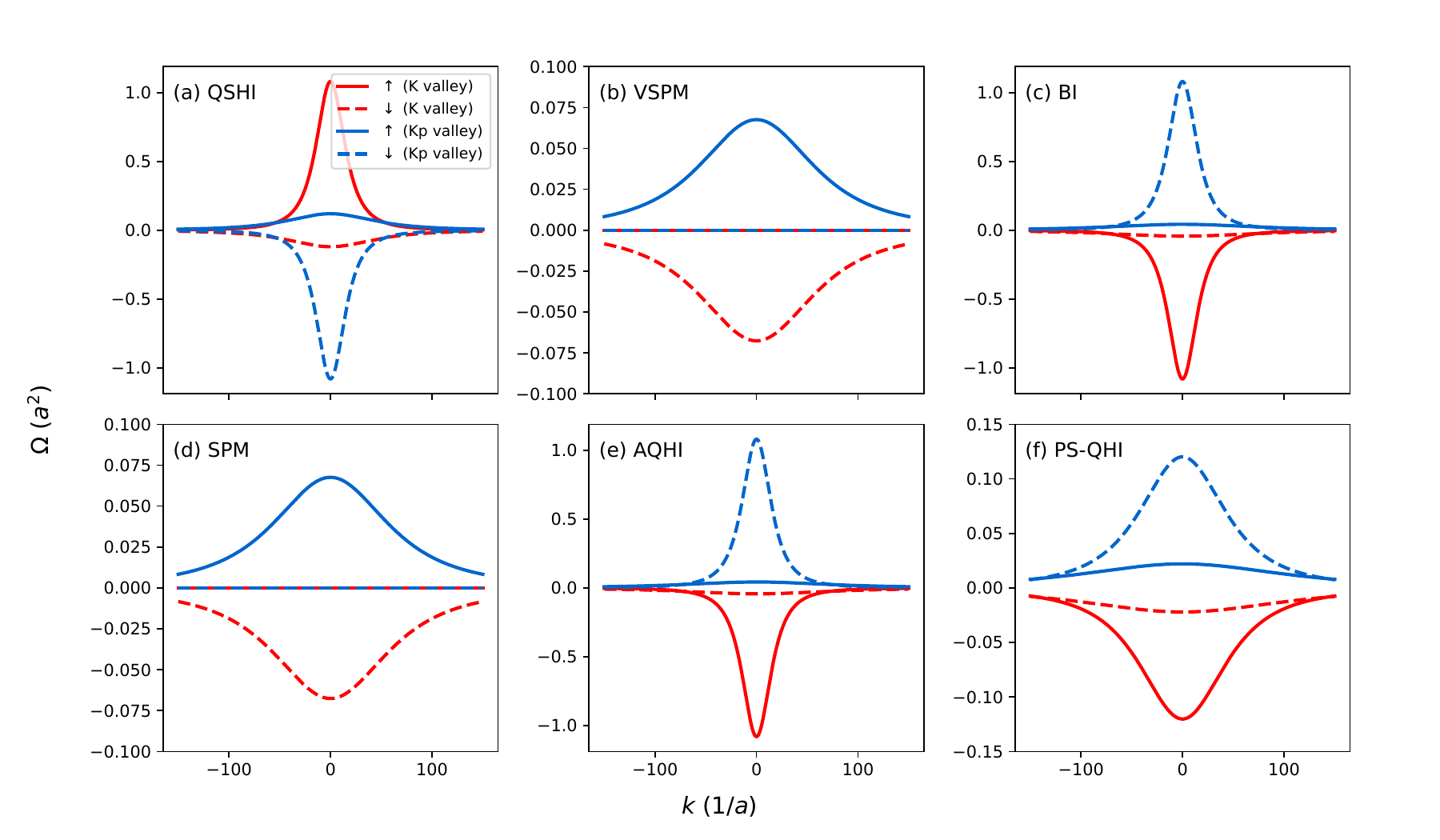}
\caption{The Barry curvature of the modified-Haldane model material along high-symmetry points in distinct topological phases at K and $K^{'}$ valleys including spin, valley, and Floquet-engineered (time-periodic) symmetry-breaking term. (a) Topological, (b) Trivial, (c) QSHI ($M=0.5\lambda_{\text{so}}$, $\lambda_{\omega}=0$), (d) VSPM  ($M=\lambda_{\text{so}}$, $\lambda_{\omega}=0$), (e) BI ($M=1.5\lambda_{\text{so}}$,  $\lambda_{\omega}=0$), (f) SPM ($M=0$,  $\lambda_{\omega}=\lambda_{\text{so}}$), (g) AQHI ($M=0$, $\lambda_{\omega}=1.5\lambda_{\text{so}}$) and (h) PS-QHI ($M=\lambda_{so}$,  $\lambda_{\omega}=1.5\lambda_{\text{so}}$) respectively. The blue, orange, and green colors are for the $N_c$, $N_s$, and $N_v$, respectively.}
	\label{fig:berry} 
\end{figure*}

\begin{figure*}[th!]
	\centering \includegraphics[width=1.0\linewidth]{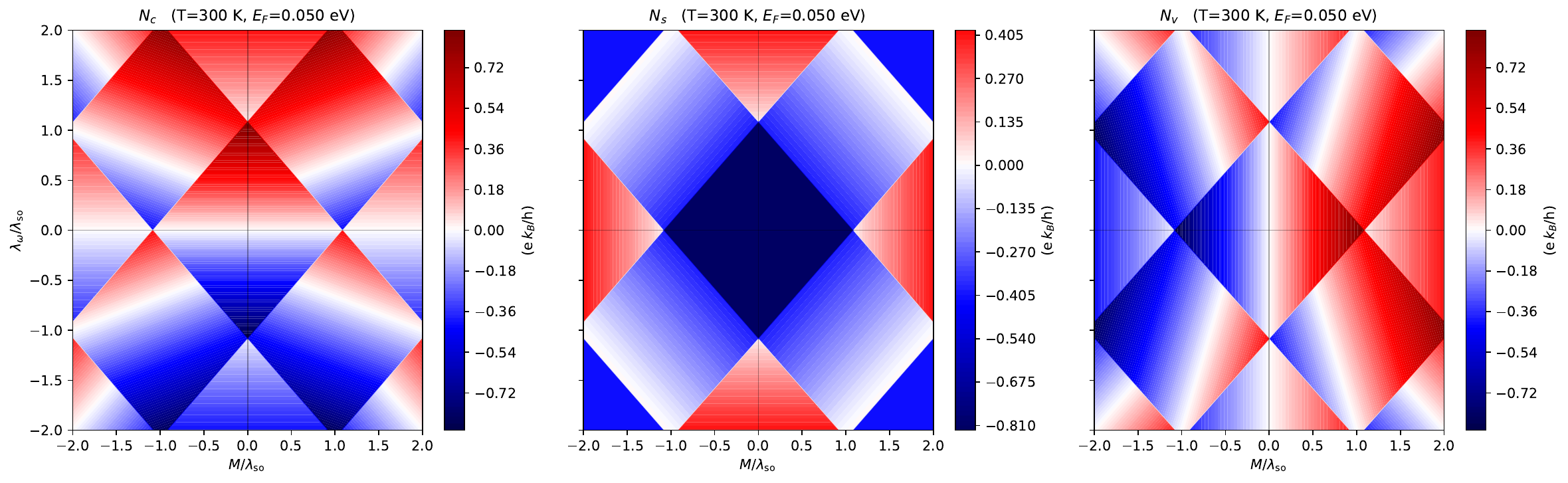}
\caption{The
charge, spin, and valley Nernst conductivities in distinct topological phases at the $K$ and $K^{'}$ valleys.   (a) QSHI ($M=0.5\lambda_{\text{so}}$, $\lambda_{\omega}=0$), (b) VSPM  ($M=\lambda_{\text{so}}$, $\lambda_{\omega}=0$), (c) BI ($M=1.5\lambda_{\text{so}}$,  $\lambda_{\omega}=0$), (d) SPM ($M=0$,  $\lambda_{\omega}=\lambda_{\text{so}}$), (e) AQHI ($M=0$, $\lambda_{\omega}=1.5\lambda_{\text{so}}$) and (f) PS-QHI ($M=\lambda_{so}$,  $\lambda_{\omega}=1.5\lambda_{\text{so}}$) respectively The blue and red colors show the corresponding changes in the value $N_c$, $N_s$, and $N_v$ in different topological states shown in [Fig.\ref{Energy}].}
		\label{fig3}
\end{figure*}

\begin{figure*}[th!]
	\centering \includegraphics[width=1.0\linewidth]{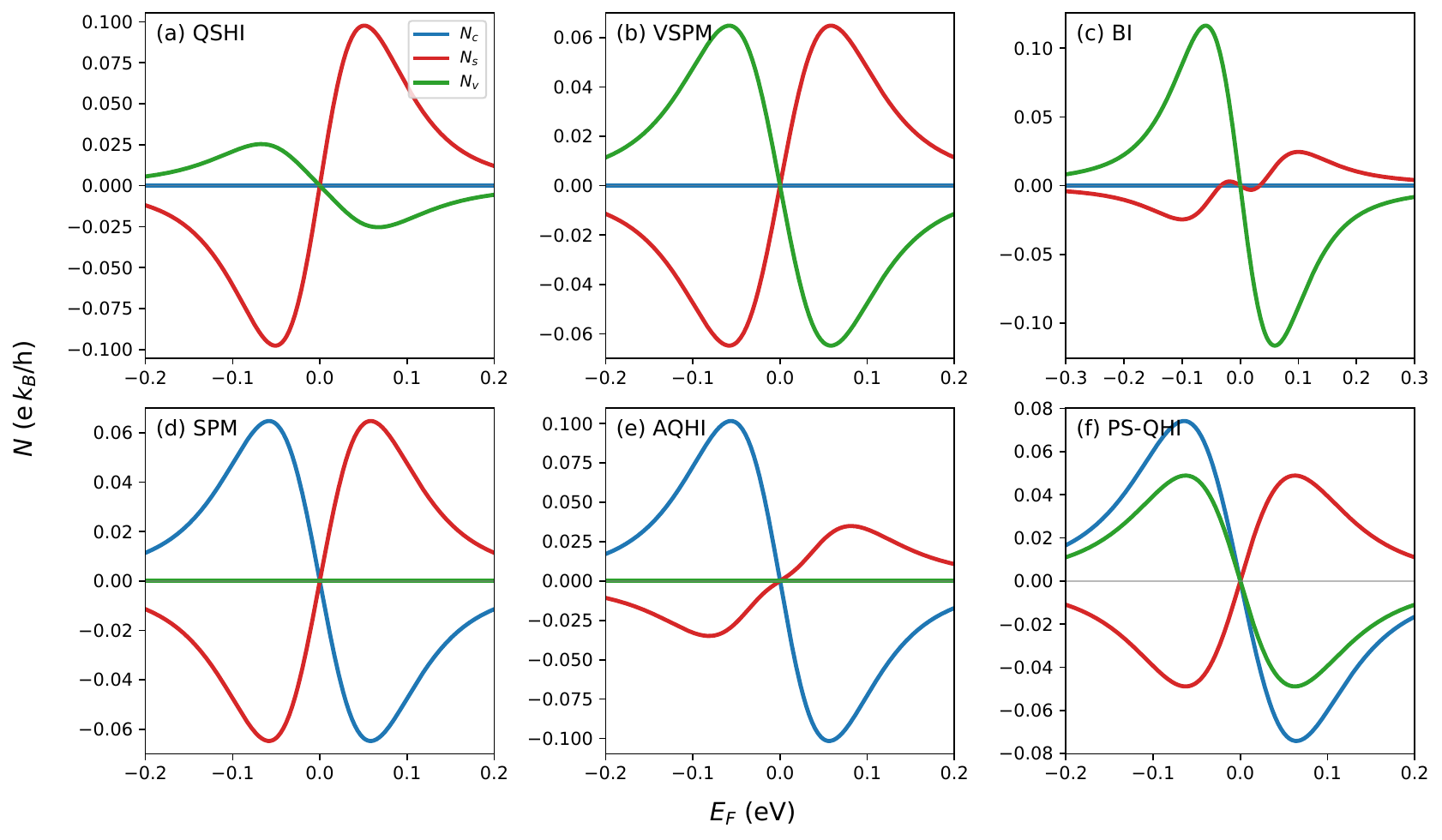}
\caption{The
charge, spin, and valley Nernst conductivities in distinct topological phases at the $K$ and $K^{'}$ valleys.  (a) QSHI ($M=0.5\lambda_{\text{so}}$, $\lambda_{\omega}=0$), (b) VSPM  ($M=\lambda_{\text{so}}$, $\lambda_{\omega}=0$), (c) BI ($M=1.5\lambda_{\text{so}}$,  $\lambda_{\omega}=0$), (d) SPM ($M=0$,  $\lambda_{\omega}=\lambda_{\text{so}}$), (e) AQHI ($M=0$, $\lambda_{\omega}=1.5\lambda_{\text{so}}$) and (f) PS-QHI ($M=\lambda_{so}$,  $\lambda_{\omega}=1.5\lambda_{\text{so}}$) respectively. The blue, orange, and green colors are for the $N_c$, $N_s$, and $N_v$, respectively.}
		\label{fig:ane_phases}
\end{figure*}

\begin{figure*}[ht!]
	\centering \includegraphics[width=1\linewidth]{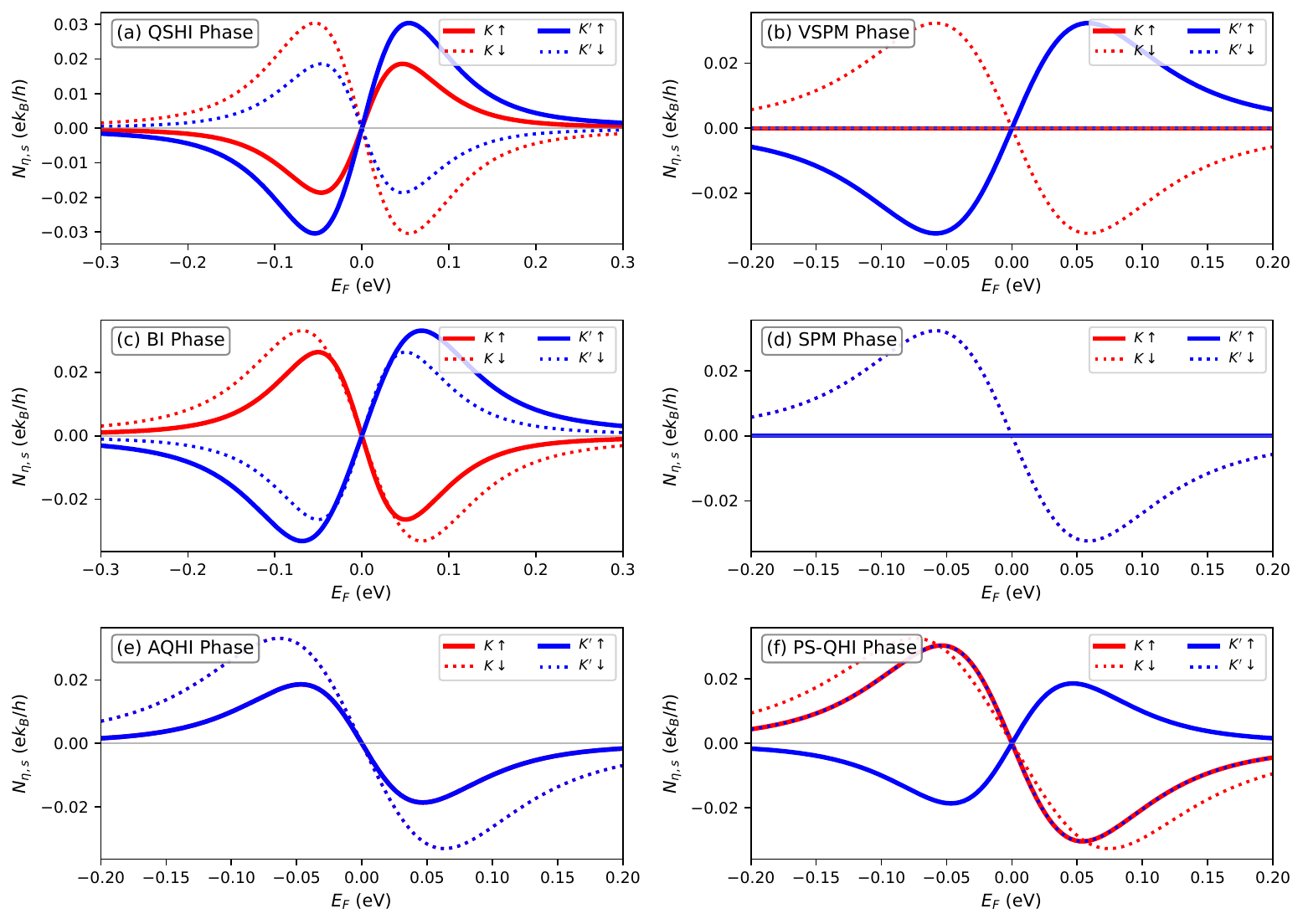}
\caption{The anomalous Nernst conductivity of the modified-Haldane Model material including spin, valley, and Floquet-engineered (time-periodic) symmetry-breaking terms in distinct topological phases at the $K$ and $K^{'}$ valleys. (a) Topological, (b) Trivial, (c) QSHI ($M=0.5\lambda_{\text{so}}$, $\lambda_{\omega}=0$), (d) VSPM  ($M=\lambda_{\text{so}}$, $\lambda_{\omega}=0$), (e) BI ($M=1.5\lambda_{\text{so}}$,  $\lambda_{\omega}=0$), (f) SPM ($M=0$,  $\lambda_{\omega}=\lambda_{\text{so}}$), (g) AQHI ($M=0$, $\lambda_{\omega}=1.5\lambda_{\text{so}}$) and (h) PS-QHI ($M=\lambda_{\text{so}}$,  $\lambda_{\omega}=1.5\lambda_{\text{so}}$) respectively.}
\label{fig:ane}
\end{figure*}

\begin{figure*}[ht!]
	\centering \includegraphics[width=13.0cm]{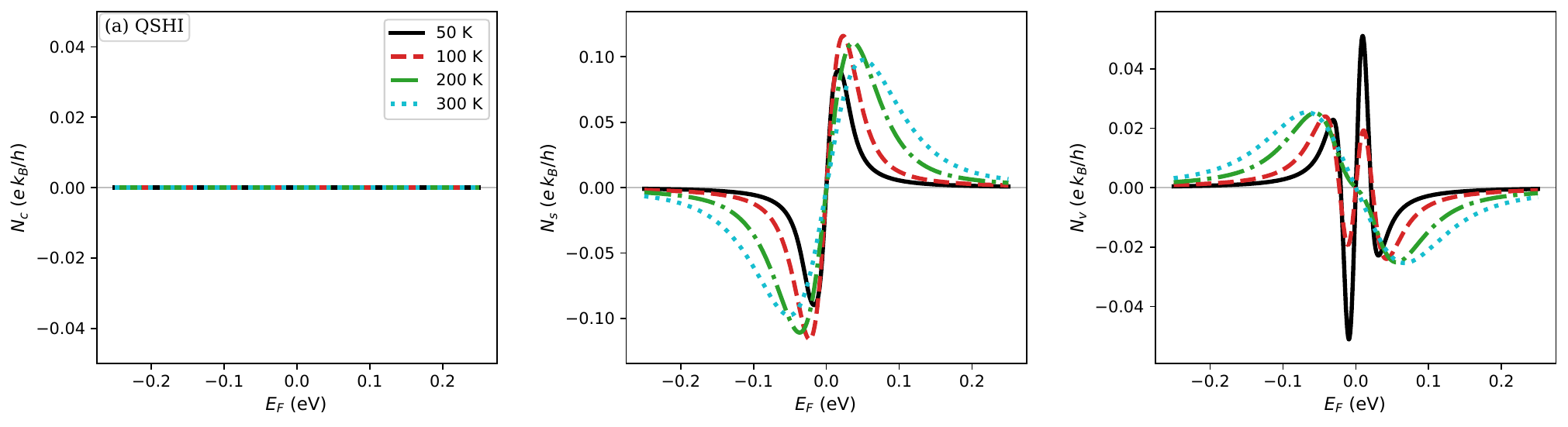}
	\centering \includegraphics[width=13.0cm]{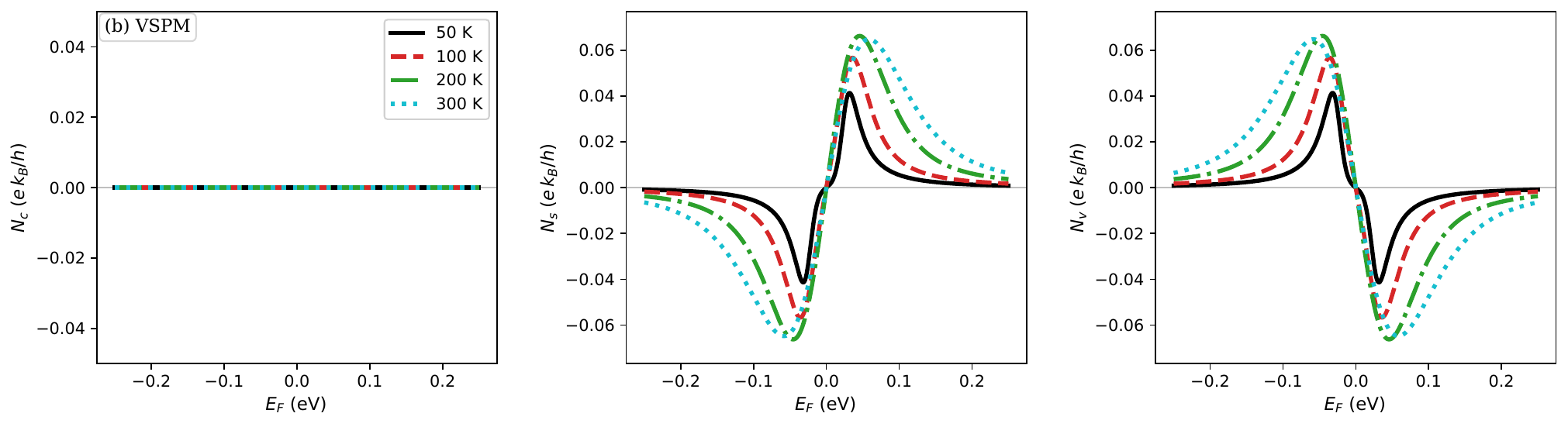}
	\centering \includegraphics[width=13.0cm]{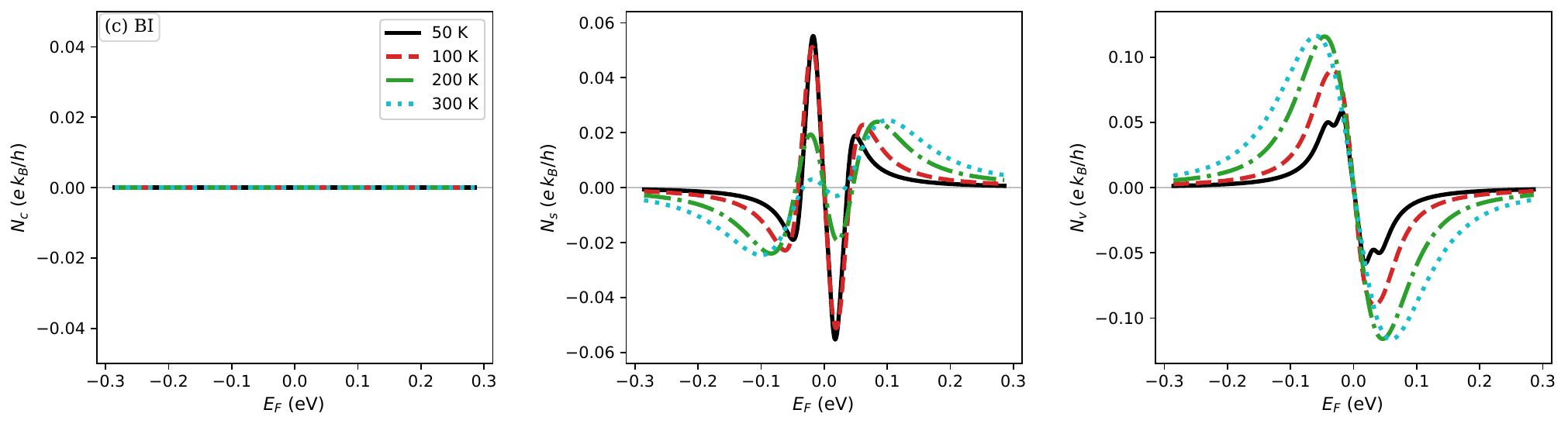}\\
    \centering \includegraphics[width=13.0cm]{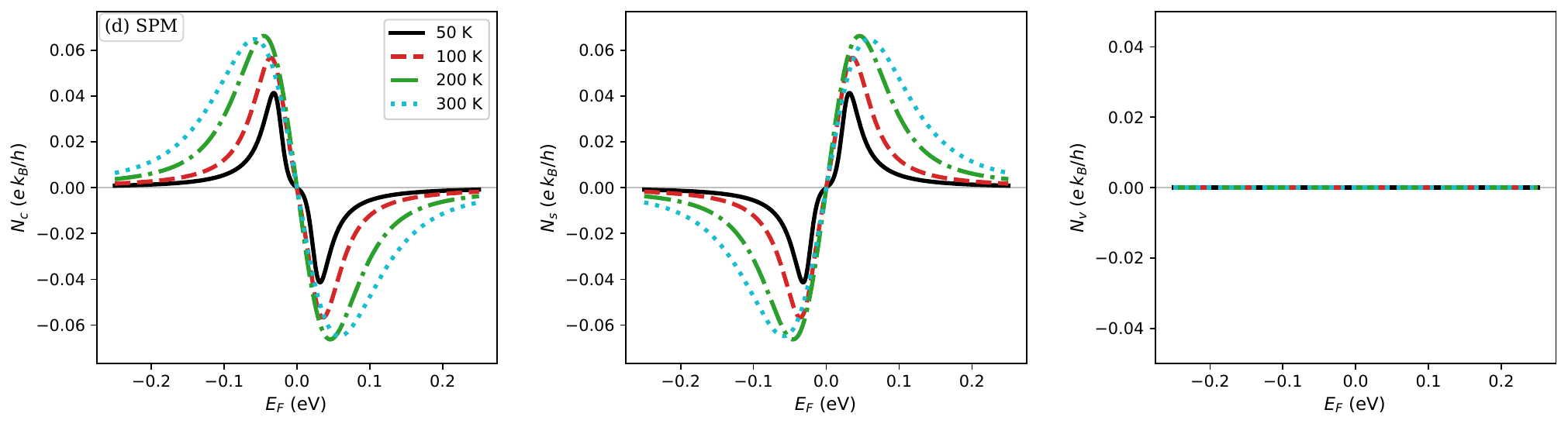}
    \centering \includegraphics[width=13.0cm]{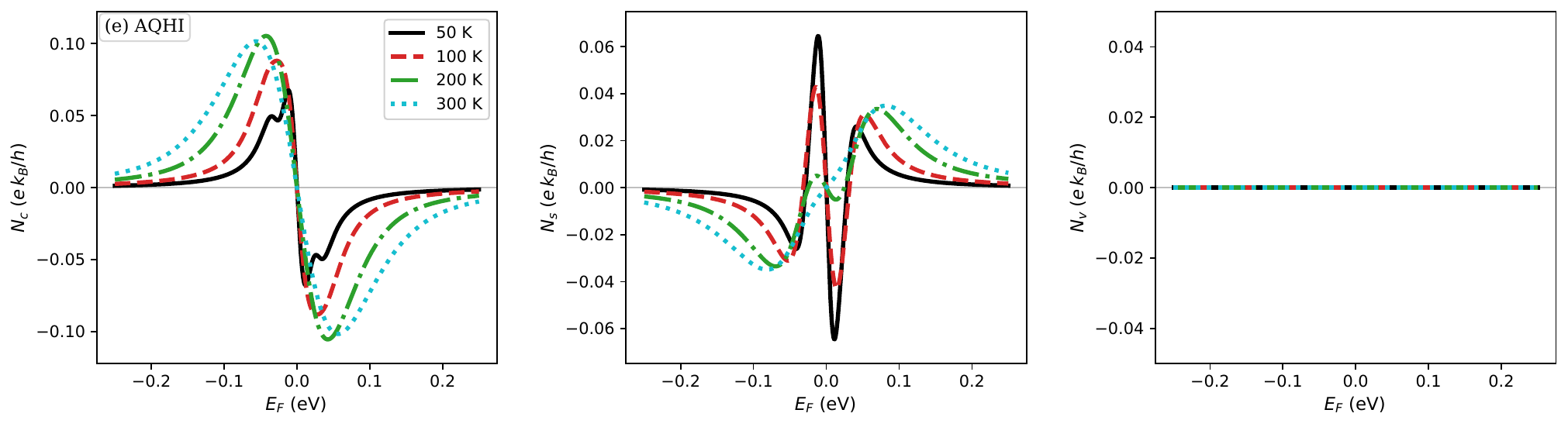}
      \centering \includegraphics[width=13.0cm]{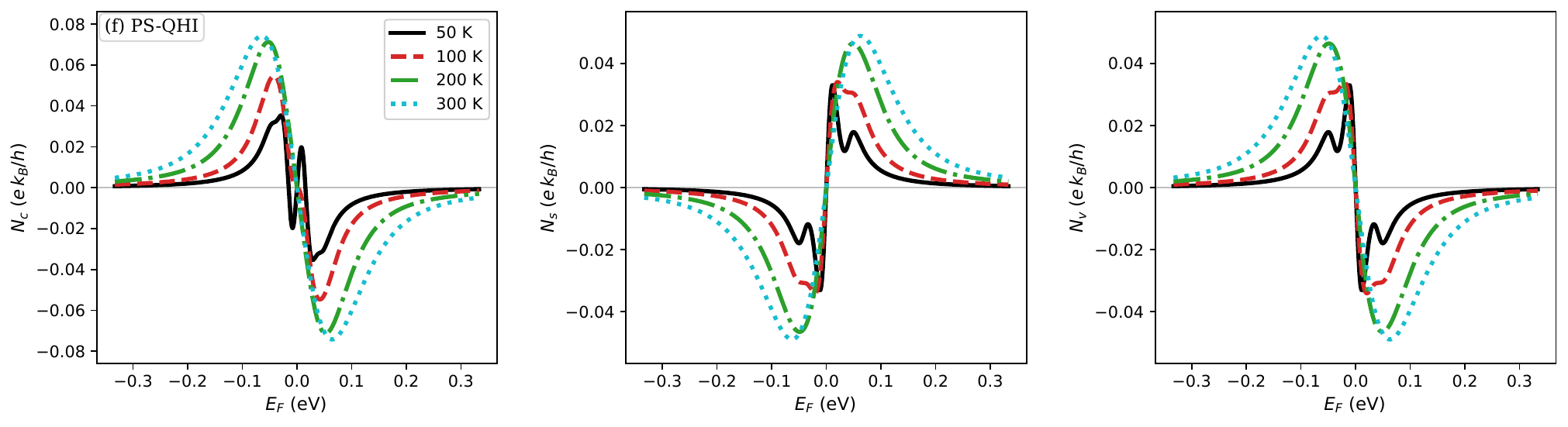}
\caption{The anomalous Nernst coefficients, $N_c$(charge), $N_s$(spin), and $N_v$(valley) vary with Fermi energy $E_F$ for temperature $T= 50\text{K}, 100\text{K},200\text{K}, 300\text{K} $ of the modified-Haldane Model material in distinct topological phases. (a) Topological, (b) Trivial, (c) QSHI ($M=0.5\lambda_{\text{so}}$, $\lambda_{\omega}=0$), (d) VSPM  ($M=\lambda_{\text{so}}$, $\lambda_{\omega}=0$), (e) BI ($M=1.5\lambda_{\text{so}}$,  $\lambda_{\omega}=0$), (f) SPM ($M=0$,  $\lambda_{\omega}=\lambda_{\text{so}}$), (g) AQHI ($M=0$, $\lambda_{\omega}=1.5\lambda_{\text{so}}$) and (h) PS-QHI ($M=\lambda_{\text{so}}$,  $\lambda_{\omega}=1.5\lambda_{\text{so}}$) respectively.}
\label{fig:nernst_vs_T}
\end{figure*}

\begin{figure*}[ht!]
	\centering \includegraphics[width=1\linewidth]{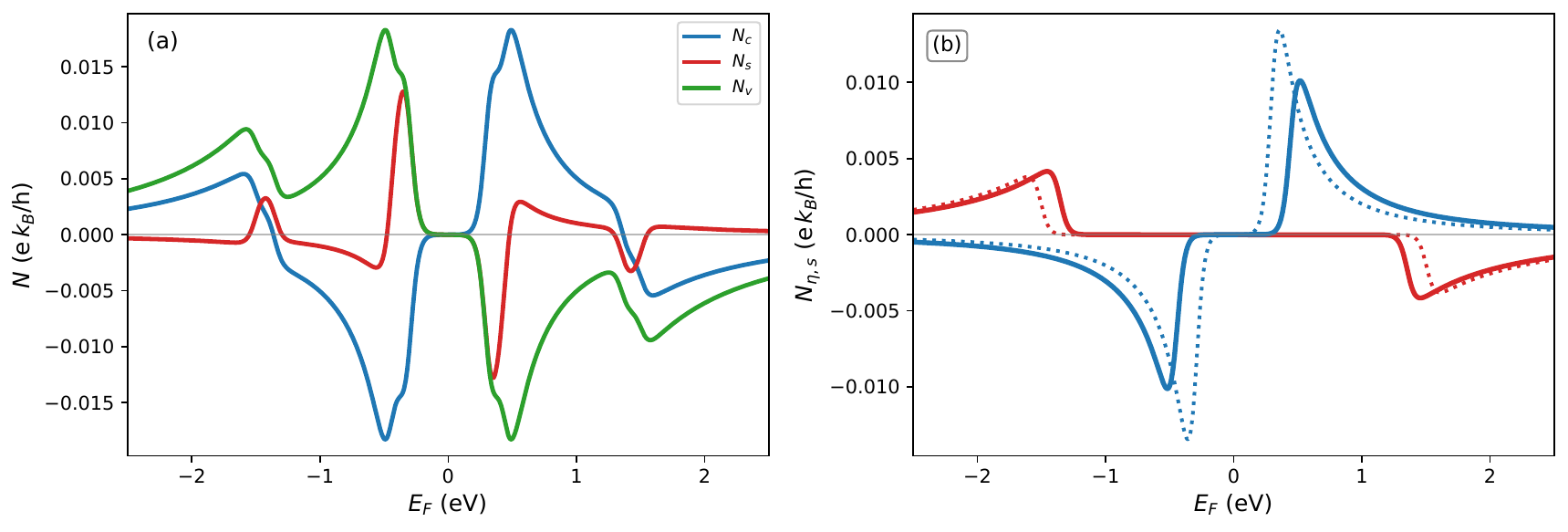}
\caption{Panel (a) illustrates the charge, spin, and valley Nernst conductivities, while panel (b) presents the spin-valley-resolved anomalous Nernst conductivity ($N_{\eta s}$), both shown as functions of the Fermi energy $E_F$ for TMDCs at the $K$ and $K^{'}$ valleys.}
\label{TMDC2}
\end{figure*}

The spin- and valley-resolved mass gap entering this expression is
$\Delta_{\mathrm{total}}=\Delta_{\tau s}-\tau s\lambda_{\mathrm{so}}+\tau\lambda_{\omega}$,
where $\lambda_{\mathrm{so}}$ denotes the intrinsic spin-orbit coupling, $M$ breaks inversion symmetry, and $\lambda_{\omega}$ represents the photoinduced time-reversal-breaking mass.
Accordingly, for $|E_F|>|\Delta_{\mathrm{total}}|$ the ANE per channel follows Eq.~(\ref{eq:N_lowT}), while it vanishes inside the gap ($|E_F|\le|\Delta_{\mathrm{total}}|$), consistent with the absence of thermally activated carriers at low temperature.
We evaluate this expression for nine representative symmetry-distinct phases including quantum spin Hall, valley-spin-polarized metallic, and anomalous quantum Hall regimes at $T=300\,\mathrm{K}$, and present both the individual channel contributions $N_{\tau s}$ and the aggregated charge ($N_c=\sum_{\tau,s}\alpha_{\tau s}$), spin ($N_s=\sum_{\tau,s}s\,\alpha_{\tau s}$), and valley ($N_v=\sum_{\tau,s}\tau\,\alpha_{\tau s}$) Nernst conductivities in units of $e k_B/h$. Although the Mott relation is strictly valid in the low-temperature limit, we use it here as a qualitative guide for larger temperatures. We also consider the system in Floquet prethermalization when systems are subjected to a high-frequency periodic driving tend to transitory states that can host intriguing physics rather than heating over extended periods of time~\cite{ho2023quantum, PhysRevB.93.155132}.

\section{Results and discussion}

We employ the light-wave-controlled Haldane framework to elucidate spin- and valley-polarized thermally driven transport in germanene and monolayer transition-metal dichalcogenides. We demonstrate that controlled variations of the Haldane parameters enable a direct evaluation of thermoelectric transport coefficients corresponding to distinct electronic phases in two-dimensional hexagonal systems. Two-dimensional Xene monolayers, where X denotes group-IV elements such as Si (silicene), Ge (germanene), Sn (stanene), and Pb (plumbene), constitute buckled analogues of graphene. A defining feature of these materials is their out-of-plane lattice distortion, which gives rise to enhanced intrinsic SOC. The SOC strengths in silicene~\cite{ezawa2012spin}, germanene~\cite{liu2011quantum}, and stanene~\cite{xu2013large} range from 1.55 7.9~meV, 24 93~meV, and up to 100~meV, respectively. Application of a perpendicular electric field $E_z$ breaks inversion symmetry and induces a staggered sublattice potential $\Delta = e l E_z$ between the $A$ and $B$ sublattices, while an off-resonant circularly polarized optical field $\lambda_\omega$ explicitly breaks time-reversal symmetry. The combined action of $E_z$ and $\lambda_\omega$ allows for valley-selective mass control of Dirac fermions, thereby driving multiple topological phase transitions~\cite{ezawa2013photoinduced,ezawa2015monolayer}.

By identifying $M=\Delta_z$, $t_2=\lambda_{\mathrm{so}}/(3\sqrt{3})$, and $\phi_s=\pm\pi/2$ in the generic Haldane Hamiltonian, the low-energy dispersion of buckled Xene monolayers is recovered as~\cite{ezawa2012spin}
\begin{equation}
E_{\tau,s,\gamma}^{\eta}(\boldsymbol{k})=
\eta\sqrt{(\hbar v_F k)^2+
\big(\Delta_{\tau s}-\tau s\lambda_{\mathrm{so}}+\tau\lambda_\omega\big)^2}.
\end{equation}

We first analyze the phase diagram and corresponding band structures of the modified Haldane model under the combined effects of sublattice asymmetry $M$ and Floquet-induced mass $\lambda_{\omega}$ as shown in Fig.~(\ref{Energy}a). These two parameters act as independent tuning knobs that control inversion and time-reversal symmetry breaking, respectively. 

In the absence of optical driving ($\lambda_{\omega}=0$), the system transitions from a quantum spin Hall insulator (QSHI) for $M<\lambda_{\mathrm{so}}$ to a trivial band insulator (BI) for $M>\lambda_{\mathrm{so}}$, with a gap closing at $M=\lambda_{\mathrm{so}}$ marking a critical point. Conversely, for $M=0$, increasing $\lambda_{\omega}$ drives the system into an anomalous quantum Hall insulating (AQHI) phase. When both perturbations are present, their interplay stabilizes intermediate phases such as spin-polarized metallic (SPM) and partially spin-polarized quantum Hall (PS-QHI) states. 

These phase transitions are governed by the sign and magnitude of the effective Dirac mass $\Delta_{\mathrm{tot}}$, which controls both the band gap and the Berry curvature distribution.


\begin{figure*}[th!]
	\centering \includegraphics[width=1\linewidth]{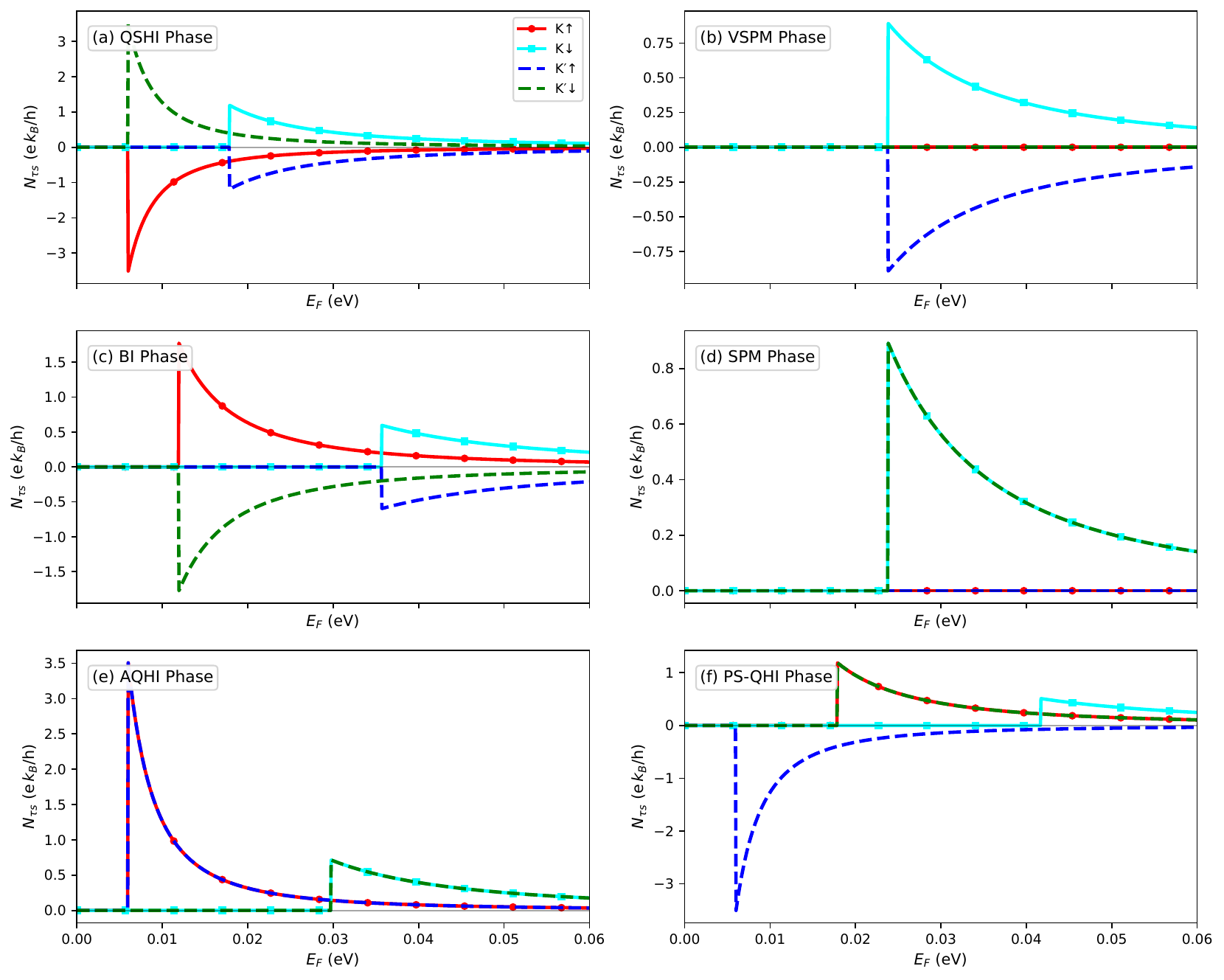}
\caption{The channel Nernst conductivity at T=300K of the modified-Haldane Model material in distinct topological phases.. (a) Topological, (b) Trivial, (c) QSHI ($M=0.5\lambda_{\text{so}}$, $\lambda_{\omega}=0$), (d) VSPM  ($M=\lambda_{\text{so}}$, $\lambda_{\omega}=0$), (e) BI ($M=1.5\lambda_{\text{so}}$,  $\lambda_{\omega}=0$), (f) SPM ($M=0$,  $\lambda_{\omega}=\lambda_{\text{so}}$), (g) AQHI ($M=0$, $\lambda_{\omega}=1.5\lambda_{\text{so}}$) and (h) PS-QHI ($M=\lambda_{\text{so}}$,  $\lambda_{\omega}=1.5\lambda_{\text{so}}$) respectively.}
\label{fig:hall_agg}
\end{figure*}


\begin{figure*}[th!]
	\centering \includegraphics[width=1\linewidth]{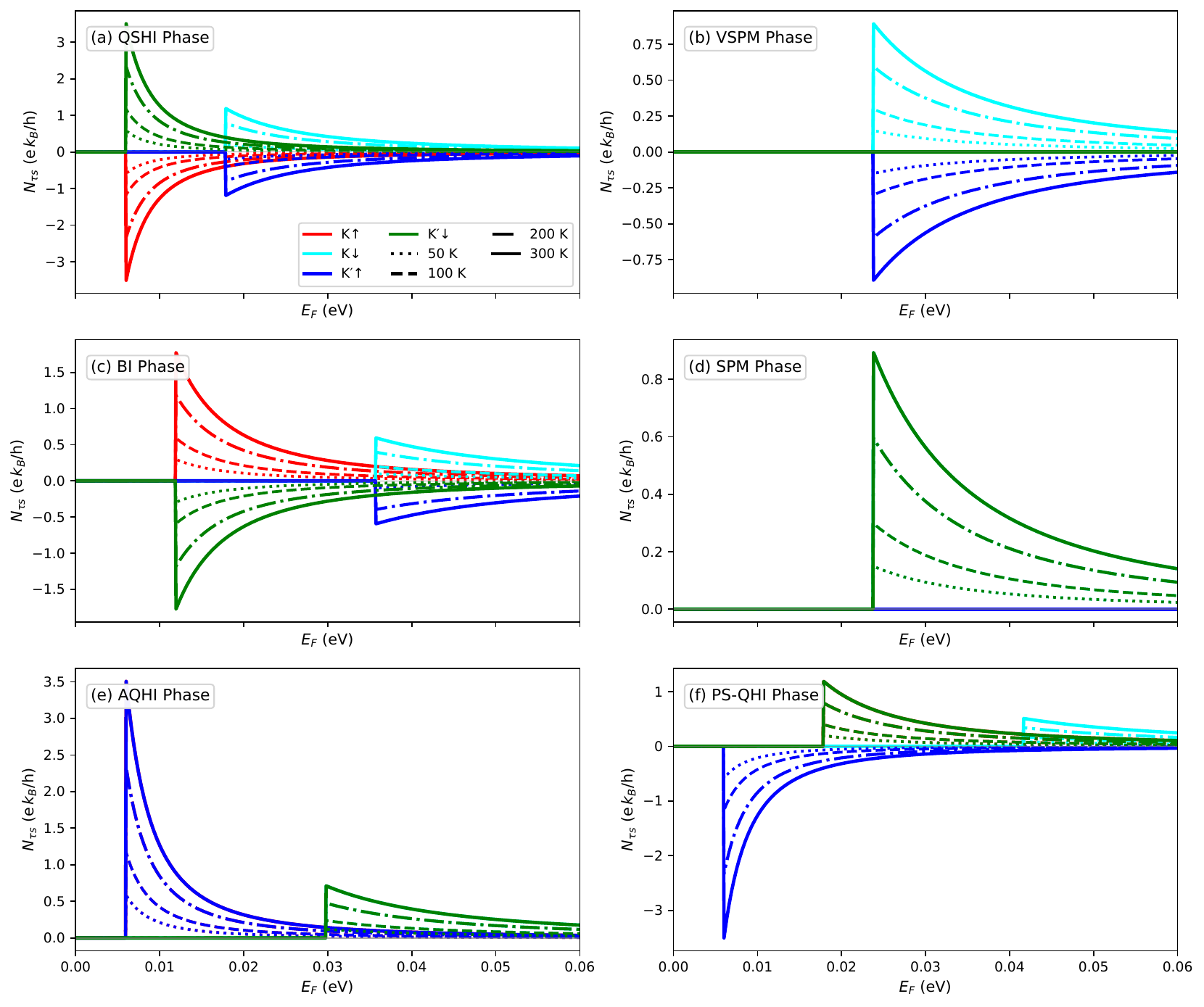}
\caption{The channel Nernst conductivity varies with Fermi energy $E_F$ for temperature $T= 50\text{K}, 100\text{K},200\text{K}, 300\text{K} $ of the modified-Haldane Model material in distinct topological phases. (a) Topological, (b) Trivial, (c) QSHI ($M=0.5\lambda_{\text{so}}$, $\lambda_{\omega}=0$), (d) VSPM  ($M=\lambda_{\text{so}}$, $\lambda_{\omega}=0$), (e) BI ($M=1.5\lambda_{\text{so}}$,  $\lambda_{\omega}=0$), (f) SPM ($M=0$,  $\lambda_{\omega}=\lambda_{\text{so}}$), (g) AQHI ($M=0$, $\lambda_{\omega}=1.5\lambda_{\text{so}}$) and (h) PS-QHI ($M=\lambda_{\text{so}}$,  $\lambda_{\omega}=1.5\lambda_{\text{so}}$) respectively.}
\label{fig:hall_chan}
\end{figure*}

For $M<\lambda_{\mathrm{so}}$, increasing either $\lambda_\omega$ or $M$ does not immediately close the bulk gap, and the system remains in the QSHI regime, as shown in Fig.~\ref{Energy}(b). When the staggered potential reaches $M=\lambda_{\mathrm{so}}$ at $\lambda_\omega=0$, the gap associated with one spin channel closes, yielding a critical point with $\Delta_{\mathrm{total}}=0$ [Fig.~\ref{Energy}(c)]. At this point, the system undergoes a transition from the QSHI phase to a spin-polarized metallic (SPM) phase with Chern numbers $(\mathcal{C}=-1,\mathcal{C}_s=1/2)$. Further increasing $M$ drives the system into a trivial band insulator (BI), as illustrated in Fig.~\ref{Energy}(d). Alternatively, setting $M=0$ and increasing the optical field strength induces a transition into the anomalous quantum Hall insulating (AQHI) phase with $(\mathcal{C}=-2,\mathcal{C}_s=0)$, as shown in Fig.~\ref{Energy}(f). At the critical point $\lambda_\omega/\lambda_{\mathrm{so}}=0.5$ and $M=0$, the bulk gap closes and the system enters the SPM phase [Fig.~\ref{Energy}(e)]. When both perturbations are applied simultaneously with $\lambda_\omega/\lambda_{\mathrm{so}}=1/2$ and $M/\lambda_{\mathrm{so}}=1/2$, the system stabilizes a polarized-spin quantum Hall insulating (PS-QHI) phase, as depicted in Fig.~\ref{Energy}(g). For sufficiently large sublattice potential ($M>\lambda_{\mathrm{so}}$), the system ultimately transitions into a trivial insulating phase with $(\mathcal{C}=0,\mathcal{C}_s=0)$. Similar topological transitions can be accessed by independently tuning either the staggered potential or the optical driving field.

The topological nature of each phase is directly revealed by the Berry curvature. The effective Dirac mass $\Delta_{\mathrm{tot}}$ determines its sign and magnitude, as illustrated in Fig.\ref{fig:berry}, with maxima concentrated close to the Dirac points.

Opposite spin and valley channels show antisymmetric Berry curvature in the QSHI phase in Fig.\ref{fig:berry}(a), which results in finite spin and valley responses but cancellation in the charge sector. A sign flip of Berry curvature occurs in the BI phase after increasing $M$ suppresses one spin-valley channel, resulting in the valley-spin-polarized metallic (VSPM) phase as shwn in Fig.\ref{fig:berry}(b). 

Further enhancement of $M$ drives the system into a trivial band-insulating (BI) phase, Fig.~\ref{fig:berry}(c), where the Berry curvature of the previously suppressed spin components reverses sign relative to the QSHI phase, while the remaining components retain their sign with slightly modified amplitude.

SPM and AQHI phases result from the asymmetric redistribution of Berry curvature between valleys caused by time-reversal symmetry breaking under finite $\lambda_{\omega}$. The Berry curvature becomes both spin- and valley-asymmetric when both $M$ and $\lambda_{\omega}$ are finite, allowing for simultaneous charge, spin, and valley transfer. This development demonstrates how Berry curvature reconstruction directly encodes topological phase transitions. The results are shown in Fig.~\ref{fig:berry}(d) and Fig.~\ref{fig:berry}(e).

Finally, when both perturbations are simultaneously present ($M=\lambda_{\mathrm{so}}$ and $\lambda_{\omega}=1.5\,\lambda_{\mathrm{so}}$), their competition stabilizes a partially spin-polarized quantum Hall insulating (PS-QHI) phase, as shown in Fig.~\ref{fig:berry}(f).

Figure~\ref{fig3} shows the charge ($N_c$), spin ($N_s$), and valley ($N_v$) Nernst conductivities as functions of $M$ and $\lambda_{\omega}$. The thermoelectric response closely follows the underlying topological phase diagram. $N_c$ to disappear in the QSHI regime, but $N_s$ and $N_v$ stay finite. On the other hand, a nonzero $N_c$ is produced by finite $\lambda_{\omega}$, indicating a violated time-reversal symmetry. Interestingly, a finite $N_v$ necessitates broken inversion symmetry, demonstrating that sublattice asymmetry governs valley Nernst transport. 
The band gap narrows, and the Berry curvature peaks sharply near phase boundaries, where the thermoelectric signal is enhanced, and the largest responses occur.


We now turn to the analysis of spin-valley-polarized thermoelectric transport in two-dimensional hexagonal systems. Throughout our numerical calculations, the temperature is fixed at $T=300\,\mathrm{K}$, otherwise it is specified. We consider a staggered sublattice potential $M$, an intrinsic spin-orbit coupling $\lambda_{\mathrm{so}}=43\,\mathrm{meV}$ appropriate for germanene, and an off-resonant right-circularly polarized light field with chirality $\gamma=+1$. The Nernst conductivities are reported in units of $e k_B/h$, while all energy scales, including the Fermi energy $E_F$, $M$, and the photoinduced mass $\lambda_\omega$, are measured in electronvolts. 

The dependence of $N_c$, $N_s$, and $N_v$ on the Fermi energy $E_F$ is shown in Fig.~\ref{fig:ane_phases}. In all phases, the Nernst response exhibits peaks near the band edges and decays at higher energies, consistent with the scaling $N \propto \Delta_{\mathrm{tot}}/E_F^2$.

Due to spin-antisymmetric Berry curvature, $N_s$ predominates in the QSHI phase, whereas $N_c \approx 0$. Asymmetry between spin-valley channels results in finite $N_s$ and $N_v$ in the VSPM and BI phases. A finite $N_c$ appears when $\lambda_{\omega} \neq 0$, signifying a violated time-reversal symmetry. 

Strong charge and spin responses are seen in the AQHI phase, but the restoration of valley symmetry results in a reduced valley signal. On the other hand, all three components are supported by the PS-QHI phase, which represents the simultaneous violation of time-reversal and inversion symmetries.
Notice that the sign of ANE is directly related to the derivative of the Hall conductivity in terms of energy.

Figure~\ref{fig:ane} presents the spin- and valley-resolved anomalous Nernst conductivity (ANC), $N_{\tau s}$, as a function of the Fermi energy $E_F$ for pristine germanene in the presence of intrinsic spin-orbit coupling, a staggered sublattice potential, and a photoinduced Haldane mass $\lambda_{\omega}$.  
The interplay of these terms drives a series of topological phase transitions, summarized in Figs.~\ref{fig:ane}(a)-\ref{fig:ane}(f).  
Based on Eq.~(\ref{eq:nernst}), we quantify the spin and valley polarizations as
$P_s=\big(|N_{\uparrow}|-|N_{\downarrow}|\big)/\big(|N_{\uparrow}|+|N_{\downarrow}|\big)$ and
$P_v=\big(|N_{K}|-|N_{K'}|\big)/\big(|N_{K}|+|N_{K'}|\big)$,
where $N_{\uparrow(\downarrow)}$ and $N_{K(K')}$ denote the spin- and valley-resolved transverse thermoelectric responses, respectively. In the QSHI phase [Fig.~\ref{fig:ane}(a)], the two spin branches carry opposite signs with identical magnitudes,
$|N_{K\uparrow}|=|N_{K'\downarrow}|$ and
$|N_{K\downarrow}|=|N_{K'\uparrow}|$.
This directly mirrors the Berry-curvature relations
$\Omega_{K\uparrow}=-\Omega_{K'\downarrow}$ and
$\Omega_{K\downarrow}=-\Omega_{K'\uparrow}$ shown in Fig.~\ref{fig:berry}(a), yielding equal and opposite spin currents consistent with preserved time-reversal symmetry and helical edge transport.  
At the valley-spin-polarized metallic (VSPM) point [Fig.~\ref{fig:ane}(b)], one spin-valley gap collapses, producing a pair of equal-and-opposite contributions (spin-down at $K$ and spin-up at $K'$), while the remaining channels partially compensate, leading to a reduced yet finite spin-valley response. When the staggered potential exceeds the SOC strength, the Berry curvature undergoes a complete sign reversal [Fig.~\ref{fig:berry}(c)], such that
$\Omega_{K\uparrow}=\Omega_{K\downarrow}<0$ and
$\Omega_{K'\uparrow}=\Omega_{K'\downarrow}>0$.
Accordingly, same-spin contributions from opposite valleys acquire opposite ANC signs with unequal amplitudes Fig.~(\ref{fig:ane}(c)].  
The diminished Berry-curvature magnitude in this band-insulating phase suppresses the overall Nernst signal, in line with the absence of a transverse charge current. Applying a right-circularly polarized off-resonant field ($\lambda_{\omega}\neq0$) explicitly breaks time-reversal symmetry and stabilizes the spin-polarized metallic (SPM) phase [Fig.~\ref{fig:ane}(d)].  
Here, the spin-up responses at opposite valleys become nearly locked, while the spin-down contributions add constructively, resulting in $P_v\simeq0$ but a finite charge Nernst conductivity. For larger $\lambda_{\omega}$, the system enters the anomalous quantum Hall insulating (AQHI) phase [Fig.~\ref{fig:ane}(e)], where the Berry curvature is identical in sign for both spins but opposite between valleys [Fig.~\ref{fig:berry}(e)].  
This produces four distinct $N_{\tau s}$ branches and a nonzero transverse charge response, with the near-quantized structure signaling the formation of a photoinduced Chern insulator. When the staggered potential and photoinduced mass coexist, the partially spin-polarized quantum Hall insulating (PS-QHI) phase emerges [Fig.~\ref{fig:ane}(f)]. In this regime, the spin-up response at $K$ and the spin-down response at $K'$ retain the AQHI-like sign structure, while the spin-up contribution at $K'$ reverses. The resulting imbalance generates a finite spin-valley Nernst current, reflecting the simultaneous breaking of inversion and time-reversal symmetries. Across all phases, the symmetry, sign, and magnitude of $N_{\tau s}$ closely track the evolution of the Berry curvature, establishing the anomalous Nernst effect as a powerful thermodynamic probe of topological order in two-dimensional Dirac materials such as germanene.

We next examine the temperature dependence of the anomalous Nernst conductivity (ANC).  
Figure~\ref{fig:nernst_vs_T} shows the charge-, spin-, and valley-resolved Nernst conductivities as functions of the Fermi energy for different topological phases and temperatures, using the same set of external parameters as in Fig.~\ref{fig:ane_phases}.  
Since the ANC is an odd function of $E_F$, contributions from both valleys are explicitly displayed. Across all phases, a clear and systematic trend emerges: increasing temperature strongly enhances the magnitude of the Nernst response, while the positions of the zero crossings remain essentially fixed. As evident from Fig.~\ref{fig:nernst_vs_T}(a)-(f), (i) in the QSHI, VSPM, and BI phases, the spin and valley Nernst components are particularly sensitive to thermal amplification;  
(ii) in the SPM and AQHI phases, the temperature dependence is dominated by the charge and spin channels; and (iii) in the PS QHI phase, all three components grow concurrently with temperature.  
The robustness of the sign-reversal points reflects the fact that the Berry-curvature topology underlying each phase is unaffected by thermal fluctuations. Although finite temperature broadens the Fermi distribution and enhances the weight of thermally activated carriers, it does not alter the band topology that governs the Berry curvature.  
Accordingly, temperature primarily rescales the overall magnitude of the ANC without shifting its characteristic features, underscoring the topological origin of the anomalous Nernst response. We provide more details regarding the Nernst conductivity behavior and its temperature dependence in the Appendix.


By choosing $\mathcal{M}=\Delta / 2$, $t_2=\Delta_{\mathrm{TMD}} / 3 \sqrt{3}$, and $\phi=+5 \pi / 6(-\pi / 6)$, we can write the dispersion relation of monolayer $M X_2$ materials~\cite{CHOI2017116,manzeli20172d} at the $K$ and $K^{\prime}$ valleys and spins as $\mathcal{E}_\eta^{\tau, s}(\boldsymbol{k})=\tau s \Delta_{\mathrm{TMD}} / 2+ \eta \sqrt{\left(\hbar v_F \boldsymbol{k}\right)^2+\left(\Delta_\tau^s\right)^2}$ where, $2\Delta_\tau^s=\Delta- \tau s \Delta_{\mathrm{TMD}}$ is the energy gap of the monolayer $M X_2$ materials

The strong spin-orbit interaction induces a pronounced spin splitting in the valence bands within each valley. At the $K$ valley, the upper valence band predominantly hosts spin-up electrons, while the lower branch is occupied by spin-down electrons. Owing to time-reversal symmetry, this spin ordering is reversed at the $K'$ valley, resulting in a valley-contrasting spin texture.

Figure~\ref{TMDC2} displays the charge ($N_c$), spin ($N_s$), and valley ($N_v$) Nernst conductivities as functions of the Fermi energy for each spin and valley sector in a monolayer TMDC.  
Owing to the large intrinsic band gap and strong spin-orbit coupling characteristic of TMDCs, the magnitude of the anomalous Nernst coefficients remains finite but relatively modest, consistent with theoretical expectations.  
Within a given valley, the two spin components contribute with opposite signs, reflecting the opposite Berry curvatures for $E_F<0$ and $E_F>0$.  
Because the states $-K\downarrow(\uparrow)$ and $K'\downarrow(\uparrow)$ are energetically degenerate, their Nernst contributions have equal magnitudes and opposite signs.  
As a result, when the Fermi level lies in the valence band, charge currents cancel, while spin- and valley-polarized transverse currents add constructively.  
This mechanism enables the generation of pure spin and valley Nernst currents driven solely by a thermal gradient, highlighting monolayer TMDCs as promising platforms for spin-caloritronic and valleytronic functionalities~\cite{PhysRevLett.115.246601}.

\section{Details of the Nernst conductivity}\label{detaild}

The validity of the Mott relation in Eq.~(\ref{eq:N_lowT}) for our multiband Dirac system is assessed by comparison with full numerical results.
Figure \ref{fig:hall_agg} display the aggregated and spin-valley-resolved Nernst conductivities across several topological phases.
The effective mass
\[
\Delta_{\mathrm{total}}
= \Delta_{\tau s}-\tau s\lambda_{\mathrm{so}}+M+\tau\lambda_{\omega}
\]
encodes the symmetry properties of the system: $\lambda_{\mathrm{so}}$ preserves both time-reversal and inversion symmetries, $M$ breaks inversion symmetry, and $\lambda_{\omega}$ acts as a valley-Zeeman term that breaks time-reversal symmetry. By tuning $M/\lambda_{\mathrm{so}}$ and $\lambda_{\omega}/\lambda_{\mathrm{so}}$, the model spans all symmetry-distinct phases, including the QSHI, AQHI, VSPM, and BI regimes. Although the ANE is not quantized and vanishes within the bulk gap at low temperature, its magnitude and sign are governed directly by $\Delta_{\tau s}$, reflecting the Berry curvature of massive Dirac bands. The valley-antisymmetric factor $\propto\tau\Delta_{\tau s}$ captures the opposite Berry-curvature monopoles at $K$ and $K'$, while the spin dependence encodes spin-momentum locking and potential spin-polarized transport.
Consequently, the charge-, spin-, and valley-resolved Nernst conductivities exhibit a universal behavior across all phases: a pronounced maximum near the band edge $E_F\simeq|\Delta_{\tau s}|$, followed by a decay $\propto E_F^{-2}$ at higher doping. This form follows directly from Eqs.~(\ref{eq:mott}) and~(\ref{eq:hall}), since differentiation yields
$\alpha_{\tau s}\propto\tau\Delta_{\mathrm{total}}/E_F^2$.
The absence of a signal for $E_F<|\Delta_{\mathrm{total}}|$ further confirms the gapped nature of the ground state and the dominance of diffusive thermoelectric transport.

We analyze the explicit temperature dependence of the ANE.
Figure \ref{fig:hall_chan} presents the charge-, spin-, and valley-resolved Nernst conductivities and the corresponding spin-valley channels as functions of $E_F$ at $T=50$, $100$, $200$, and $300$~K for several topological phases, using the same external parameters as in Fig.~\ref{fig:ane_phases}.
Across all regimes QSHI, VSPM, BI, SPM, AQHI, and PS-QHI the magnitude of the Nernst response increases systematically with temperature, while the locations of peaks and sign changes remain essentially unchanged.
This robustness demonstrates that thermal broadening primarily rescales the Berry-curvature-driven thermoelectric response without altering the underlying topological structure of the bands.

\section{Conclusion}
In conclusion, using a modified Haldane framework that includes intrinsic spin-orbit coupling, a staggered sublattice potential, and Floquet-induced time-reversal symmetry breaking, we have studied the anomalous Nernst effect in two-dimensional Dirac systems. We can systematically resolve charge-, spin-, and valley-dependent thermoelectric responses across several topological phases with this unified technique.
We find that the symmetry characteristics of the system and the corresponding Berry-curvature distribution dictate the nature of the Nernst response. Specifically, a nonvanishing valley Nernst response appears only when inversion symmetry is broken in combination with Floquet driving, whereas a finite charge Nernst conductivity requires broken time-reversal symmetry. The growth of the effective Dirac mass and the associated redistribution of Berry curvature in momentum space are closely related to the phase-dependent behavior of the Nernst coefficients.

An effective external tuning parameter for regulating the thermoelectric response's sign and magnitude is provided by the Floquet-induced mass term. The Nernst conductivities show distinctive peaks close to the band edges and sign shifts linked to topological phase transitions in all phases, and their total magnitude rises with temperature without changing the underlying structure.

Strong intrinsic spin-orbit coupling and spin-valley locking produce comparable spin- and valley-polarized thermoelectric responses when extended to monolayer transition-metal dichalcogenides, demonstrating that the suggested process is not limited to a particular material platform.
These findings give a consistent framework for studying spin- and valley-resolved Nernst effects in a wide class of two-dimensional materials and reveal a direct relationship between symmetry breaking, Berry curvature, and thermoelectric transport in Floquet-engineered Dirac systems.

\section*{Acknowledgement}
I.K. and G.X. acknowledge the financial support from the NSFC under grant No. 12174346.

\section*{References}
\bibliographystyle{apsrev4-2.bst}
\bibliography{refr}
\end{document}